# On the critical character of plasticity in metallic single crystals

Thiebaud Richeton [a], Patrik Dobron [b], Frantisek Chmelik [b],
Jérôme Weiss [a] and François Louchet [a]

[a] Laboratoire de Glaciologie et Géophysique de l'Environnement, CNRS
54 rue Molière, BP 96, 38402 St Martin d'Hères Cedex, France
[b] Department of Metal Physics, Charles University,
Ke Karlovu 5, CZ-121 16 Prague 2, Czech Republic

**Abstract:**

Previous acoustic emission (AE) experiments on ice single crystals, as well as numerical simulations, called for the possible occurrence of self-organized criticality (SOC) in collective dislocation dynamics during plastic deformation. Here, we report AE experiments on hcp metallic single crystals. Dislocation avalanches in relation with slip and twinning are identified with the only sources of AE. Both types of processes exhibit a strong intermittent character. The AE waveforms of slip and twinning events seem to be different, but from the point of view of the AE event energy distributions, no distinction is possible. The distributions always follow a power law given by $P(E) \sim E^{-t_E}$, with $t_E = 1.5 \pm 0.1$, even when multi-slip and forest hardening occur. The exponent $t_E$ is in perfect agreement with those previously found in ice single crystals. Along with observed time clustering and interactions between avalanches, these results are new and strong arguments in favour of a general, SOC-type, framework for crystalline plasticity.

## Introduction

In the past few years, it has become obvious that the processes that underlie plastic deformation are far from being smooth and steady. In crystalline materials with high dislocation mobility, there is now growing evidence that such processes are governed by intermittency and avalanches of dislocations moving through the material [1]. Acoustic emission (AE) experiments on creeping ice as well as numerical simulations [2] indicate that during plastic deformation, the dislocations self-organize into a scale-free pattern of dislocation avalanches, characterised by power law distribution of avalanche amplitudes $A_0$, i.e. $P(A_0) \sim A_0^{-t_A}$. AE stems from elastic waves generated by sudden irreversible structure changes within the material, like cooperative dislocation motion. It is thus a very powerful tool for exploring plastic deformation.

Since the pioneering work of J. Kaiser [3], AE has frequently been used to investigate plastic deformation of materials, notably metals. However, there has been only little evidence on AE from metallic single crystals [4,5]. Moreover, these experiments did not focus on the statistical properties of the AE events. According to the authors' knowledge, the AE experiments on ice crystals mentioned above are the only ones in which such a statistical analysis of the AE events occurring during plastic deformation was performed. In addition to power law size distributions, dislocation avalanches in ice single crystals were characterized by strong intermittency [2], time correlations and aftershock triggering [6], as well as fractal distributions of avalanche locations and complex space-time coupling [7]. These results were found to be independent of applied stress [2] and temperature [8]. Moreover, they are in agreement with discrete dislocation dynamics (DDD) simulations [2] and phase-field models [9]. Consequently, they suggest a general framework for dislocation-related plasticity. The concept of self-organized criticality (SOC) [10] might apply to collective dislocation dynamics in single crystals. The occurrence of SOC in a system does not depend on microscopic details (such as, e.g., the nature of atomic bonds) related to the individual behaviour of the interacting entities [11]. According to the universal character of SOC, the observed scale-free pattern should not be specific of ice and may emerge during plastic deformation of other materials as well.

In the present paper, AE experiments on single crystals of hexagonal metals (cadmium and zinc) are reported. These materials differ of ice, which has some specific properties in terms of plastic deformation. Actually, ice is an hcp material presenting a very strong plastic anisotropy [12]. Consequently, plastic deformation of ice single crystals occurs almost exclusively by basal glide. There is no (forest) hardening and no twinning. In Cd and Zn, the plastic anisotropy is not as strong. Depending on the sample orientation with respect to the tensile or compression axis, glide on nonbasal planes and twinning are possible. As compared to metals, ice has a low elastic modulus (E ~ 9 GPa) and the intermolecular forces stem from hydrogen bonds. Moreover, for practical reasons, the quoted AE experiments on ice were always performed under a constant applied stress (compression creep experiments) and at a temperature T close to the melting point $T_m$ ($T/T_m$ ~ 0.95). In the present work, tensile tests are performed at constant crosshead speed and at room temperature ($T/T_m$ ~ 0.4 for both Cd and Zn). The goal of the present paper is to analyse the AE parameters recorded during plastic deformation of Cd and Zn single crystals with respect to the results obtained on ice single crystals. Particularly, we will attempt to understand to what extent nonbasal glide, dislocation-hardening and twinning could influence the global dislocation dynamics.

**Experimental procedure**

Two different hexagonal materials were tested: Cd single crystals (of commercial purity) and Zn-0.08%Al single crystals. Single crystals of Cd and Zn were grown under an argon atmosphere from raw metals by the horizontal Bridgman technique, using a graphite mould and oriented seed crystals. The single crystal orientations were determined by the back-reflection X-ray Laue technique. Table 1 summarizes the main characteristics of the samples tested. *?* is the initial angle between the c-axis and the tensile axis and *μ* is the initial angle between the slip direction and the tensile axis. The different orientations yield values of the Schmid factor ranging from 0.14 to 0.48 for Cd and from 0.33 to 0.46 for Zn-0.08%Al. The specimens were deformed in a computer-controlled testing machine in tension, at room temperature and at a constant crosshead speed, giving initial strain rates from $1.11 \times 10^{-3}$ $s^{-1}$ to $1.96 \times 10^{-3}$ $s^{-1}$. During each test, a miniaturized MST8S transducer (producer DAKEL-ZD Rpety, Czech Republic, diameter 3 mm, frequency band 100-600 kHz) was fastened to the surface of the cylindrical specimen, with silicon grease as a coupling material. The dynamic range between the amplitude detection threshold $A_{min}$ ($3 \times 10^{-3}$ V, or 30 dB) and the maximum recordable amplitude (10 V, or 100 dB) was 70 dB, i.e. 3.5 orders of magnitude. The AE acquisition system (Euro Physical Acoustics, Mistras 2001) allocates different parameters to each detected event, such as the arrival time $t_0$ (the time at which the AE event signal exceeds first $A_{min}$), the maximum signal amplitude $A_0$, the AE signal duration ***d***, and the AE signal energy $E$ [(1)]. More detail about the AE acquisition system and the determination / signification of the AE parameters can be found in [8].

| Sample | Material | Diameter (mm) | Length (mm) | ? (°) | μ (°) | Schmid factor |
|--------|----------|---------------|-------------|-------|-------|---------------|
| a1 | Cd | 3.72 | 17 | 72 | 22 | 0.29 |
| a2 | Cd | 3.95 | 22 | 64 | 35 | 0.36 |
| a3 | Cd | 3.58 | 25 | 48 | 44 | 0.48 |
| a4 | Cd | 3.60 | 30 | 43 | 51 | 0.46 |
| a5 | Cd | 3.98 | 25 | 67 | 33 | 0.33 |
| a6 | Cd | 3.75 | 30 | 70 | 24 | 0.31 |
| a7 | Cd | 3.72 | 30 | 14 | 82 | 0.14 |
| b1 | Zn-0.08%Al | 5.39 | 25 | 57 | 33 | 0.46 |
| b2 | Zn-0.08%Al | 5.47 | 25 | 70 | 22 | 0.33 |

**Table 1. Summary of sample information.**

## Results

*Stress-strain curves*

Figure 1a shows the stress-strain curves for Cd and figure 1b for Zn-0.08%Al. Two Cd specimens (a3 and a4) and one Zn-0.08%Al specimen (b1) have a basal orientation favourable for glide (Schmid factor > 0.45) and exhibit a typical region A of easy basal glide, characterized by a low stress plateau. Following easy basal glide, all curves show a second stage B with a sharp increase of the stress, characteristic of strain hardening, due to the activation of nonbasal slip systems. Furthermore, in region C, serrations are observed prior to failure (a2, a3, a4, b1, b2) or necking (a1, a5, a6, a7). In single crystals, these serrations are twin signatures. Twinning is an important deformation mechanism in hcp metals, as the number of basal glide systems is limited. As twinning requires stress concentrations [13], the total amount of twinning is supposed to increase in the course of a test. In Cd and Zn, there is one active twinning system, {10-12} <-1011>. Unlike polycrystals where twin growth is usually limited by grain boundaries, there is no such restriction in single crystals. Consequently, twins may be very large and may result in large stress relaxations. As seen in figures 1a and 1b, where samples with a low Schmid factor exhibit the largest serrations, twinning is enhanced by an unfavourable orientation of the basal planes. Moreover, twinning is known to be an important source of AE [14]. Therefore, unlike plastic deformation of ice, plastic deformation of Cd and Zn single crystals can exhibit two sources of AE: dislocation glide avalanches and twinning. However, as a process involving fast multiplication and movement of dislocations produced by local overstress [15], twinning can also be considered as a kind of dislocation avalanche process.

---

[(1)] $E$ is calculated by the AE system as $E = \int_{t_0}^{t_0+d} A^2(t)dt$.

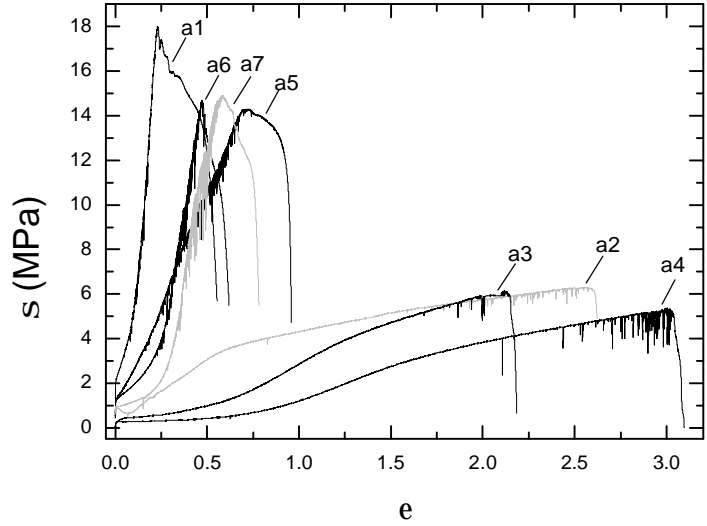

**Figure 1a. Stress-strain curves for Cd samples.**

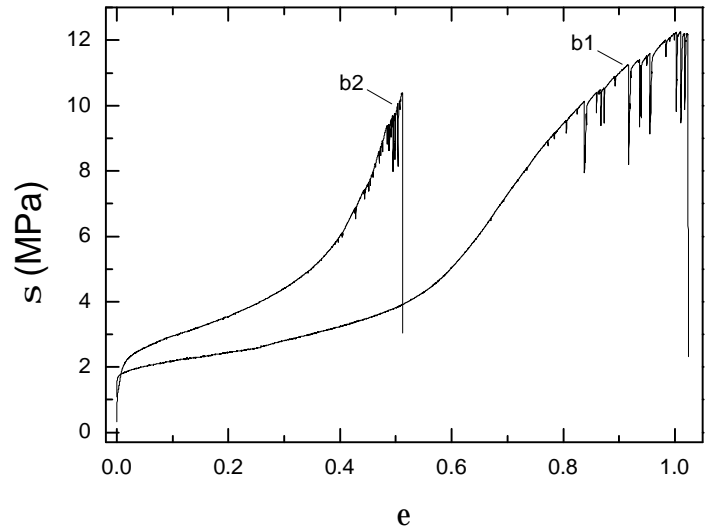

**Figure 1b. Stress-strain curves for Zn-0.08%Al samples.**

*AE activity*

An AE activity, exhibiting a strong intermittent character (Fig. 2 inset) is observed in all experiments. The intermittent AE recorded during stage A (easy basal glide) where twinning is absent, indicates the occurrence of dislocation glide avalanches, as it was observed previously in ice. Figure 2 shows the evolution of the AE activity during plastic deformation of the large Schmid factor sample b1. It can be seen that the ratio $SA_0/e$ of the AE activity (sum of all the AE amplitudes over a time window) *vs.* the experimental (axial) strain, is almost constant at the beginning of the test, which corresponds to the region of easy basal glide. This result is in agreement with the experiments performed on ice single crystals [16], which showed a direct proportionality between the AE activity and the macroscopic deformation, and with the model discussed in [8], where the maximum amplitude of an acoustic wave ($A_0$) was related to the strain dissipated by the corresponding dislocation avalanche. Such an AE source model might also be consistent with twinning, as indicated by the numerous dislocation-based models that have been proposed to interpret twinning mechanisms. For instance, the model described in [17] suggests repeated double cross-slip to a neighbouring plane, followed by the operation of a partial Frank-Read source, to produce overlapping faulted loops which define a twin band.

However, in figure 2, following the stage A of easy basal glide, a sharp increase of the ratio $\mathbf{\mathit{S}A_0/e}$ is observed in correlation with the hardening of the material and the occurrence of serrations. As a matter of fact, in presence of multi-slip and twinning, AE activity is no more proportional to the macroscopic deformation. First of all, it must be underlined that in the present case, avalanches occur on different planes (twinning is active on pyramidal planes whereas slip is active on basal and other planes), which implies different contributions of resolved strain to macroscopic, axial strain. But this point cannot explain on its own the very strong rise of $\mathbf{\mathit{S}A_0/e}$. It appears most likely that such an increase is related to the increase of the proportion of AE events originating from twinning among the whole AE event population. Accordingly, in case of a twinning event, $A_0$ might still be related to the deformation achieved by the avalanche, but then the corresponding coefficient of proportionality should be much higher than in the case of a slip event. In fact, some studies [14,18] tried to compare final twin extents (from *post-mortem* microstructural analyses) to the characteristics of associated AE signals, but found no straightforward correlation. Hence, in case of twinning, the relation between deformation and AE is not clear and remains an open question. Thus, in the present experiments, it does not seem appropriate to interpret the AE generated by a plastic instability in terms of dissipated deformation. Instead an interpretation in terms of dissipated energy may be more sound. Indeed, in the study of dynamical systems, dissipated energy is usually used as a universal and relevant parameter to characterize the size of an instability. Moreover, for AE related to microcracking [19,20], reasonable arguments support the assumption that the AE energy radiated by the acoustic wave is proportional to the whole energy dissipated by the source mechanism. Here, we make a similar assumption for both slip [21] and twinning events. As a consequence, the further detailed analysis is focussed on the statistical properties of AE event energy $E$, assumed to be proportional to the energy dissipated at the AE source whatever the source mechanism.

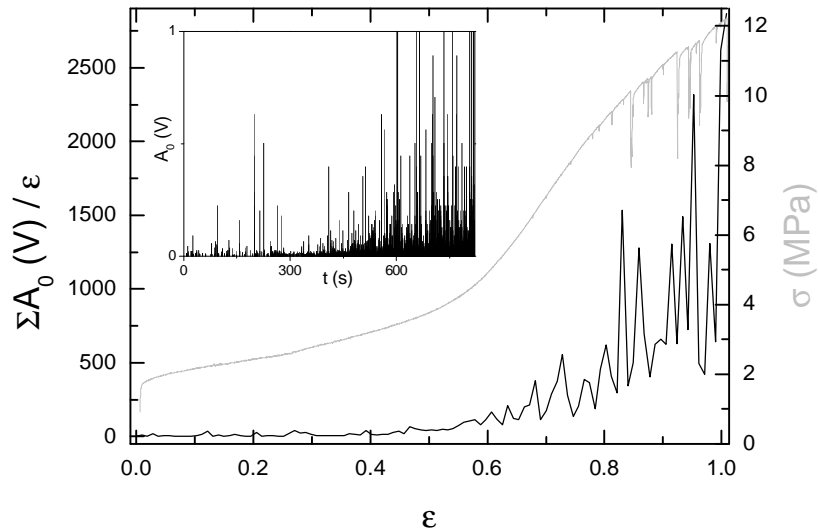

**Figure 2. Evolution of the AE activity during deformation of the sample b1. The black curve represents the amplitudes of the AE events ($A_0$) cumulated over a time window of 7s and normalized by the corresponding strain. The grey curve is the stress-strain curve of the sample b1. The inset shows the recorded AE signal.**

*AE energy distributions*

The first significant result of the present work is that the probability density function (pdf) of the AE event energies obeys a power law distribution, $P(E) \sim E^{-t_E}$, independently of the material composition and the orientation of the basal planes (Fig. 3). For the sake of clarity, figure 3 shows only the pdf of samples a3, a5 and b2 but a similar result is obtained for any of the tested samples. As a matter of fact, samples a3 and a5 (Cd) have different orientations of the basal planes, whereas the sample b2 is of different composition (Zn-0.08%Al). The measured values of the exponent $t_E$ range from 1.4 to 1.6. These values are very close to those found for dislocation avalanche size distributions in single crystals of ice [8]: $t_A = 2.0 \pm 0.1$, i.e. $t_E = 1.5 \pm 0.1$ for the distributions of the associated energies (because of the $E \sim A^2$ scaling which applies also for the present experiments (not shown)). Of course, the present distributions refer not only to dislocation avalanches related to glide but also to twinning. For this reason, we compared a distribution obtained within a region of easy basal glide where no twinning is expected with one obtained in subsequent stages (serrated stress-strain curve) where the proportion of AE from twinning is much larger (Fig. 4). Absolutely no difference in the AE energy distributions is found between these two typical regions.

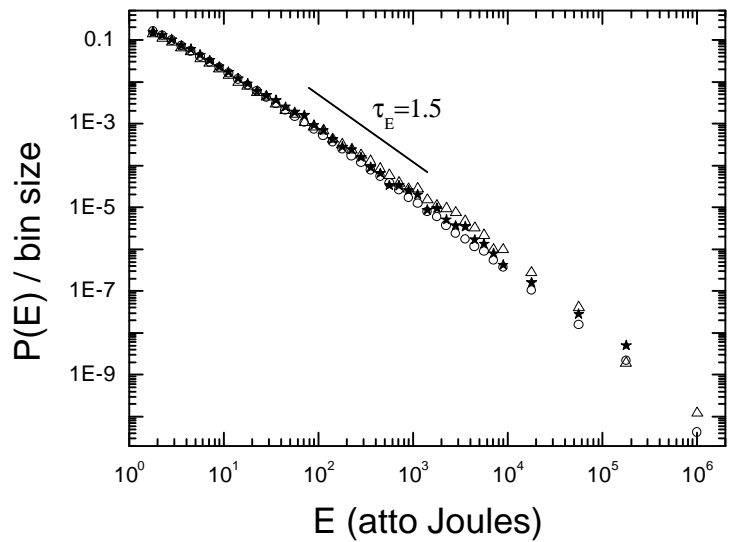

**Figure 3. Probability density functions of AE energies. Open triangles: test on sample a3. Open circles: test on sample a5. Full stars: test on sample b2.**

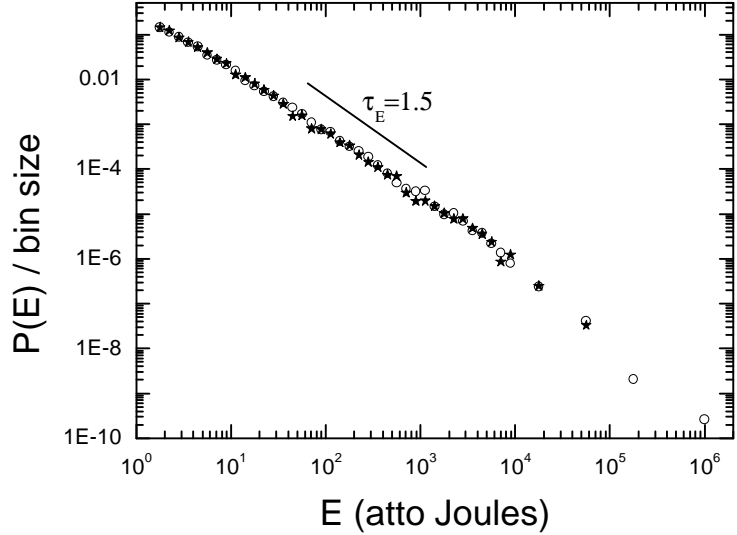

**Figure 4.** Comparison of the probability density functions of AE energies for the stage of easy basal glide and the subsequent stage with twinning of the test a3. Full stars: stage of easy basal glide (**e** < 1, see curve a3 in figure 1a). Open circles: subsequent stage with twinning (**e** > 1.6).

*Twinning and dislocation glide*

From the simple analysis of the AE energy distributions, no distinction between twinning and dislocation glide is possible. Hence, from this point of view, these phenomena appear to belong to the same global dynamics. However, a difference between AE signals corresponding to dislocation slip and to twinning was noticed by Bovenko et al. [5] during plastic deformation of Zn single crystals. In our experiments, two main types of AE waveforms can be observed (Fig. 5). According to the arguments developed in [18], type 2 pattern may be associated with a twinning process. The first high-frequency burst should correspond to the initial high-rate energy release occurring during twin initiation. As in single crystals, twins normally extend over large distances, the sequel of the AE signal should reflect the sustained process of twin growth. As reported in [18], for the same amplitude, the duration of the AE signals associated with slip is substantially smaller than the duration of the signals associated with twinning. Therefore, the type 1 pattern could be indicative of a dislocation glide avalanche, as those observed in ice [8]. In [18], the distinction between the two types of patterns was evaluated in terms of the ratio between AE maximum amplitude ($A_0$) and AE root mean square amplitude ($A_{rms}$). This parameter characterizes the distribution of signal energy over its time span. The type 2 events yield lower values of this parameter than those of type 1. Figure 6 reports the evolution of this parameter with time during the loading of the large Schmid factor sample a3. In fact, we did not record the value of $A_{rms}$ and used instead the square root of the recorded AE energy ($E^{1/2}$). Figure 6 shows that the average value of $A_0/E^{1/2}$ evolves from a high value in the region of easy basal glide to a lower and relatively constant value at the end of the test when AE from twinning becomes dominant. According to these observations, $A_0/E^{1/2}$ seems to be a good parameter to distinguish, from a statistical point of view, AE events related to dislocation glide from AE events related to twinning. Therefore, two distinct populations of AE events can be separated on the basis of a (somehow arbitrary) threshold on $A_0/E^{1/2}$. Events with low $A_0/E^{1/2}$ values will be called T-type events and events with high $A_0/E^{1/2}$ values S-type events. Of course, the real proportion of twinning in the two different groups remains unknown. We just verified that the relative proportion of T-type events evolves in a similar way as the population of AE due to twinning is expected to do. Nevertheless, this arbitrary distinction enables us to verify that the probability distribution of AE event energies remains unchanged whether the analysis is made from a population of AE events exhibiting a supposedly small (S-type) or large (T-type) proportion of twinning. Results in full agreement with those of figure 4 are obtained.

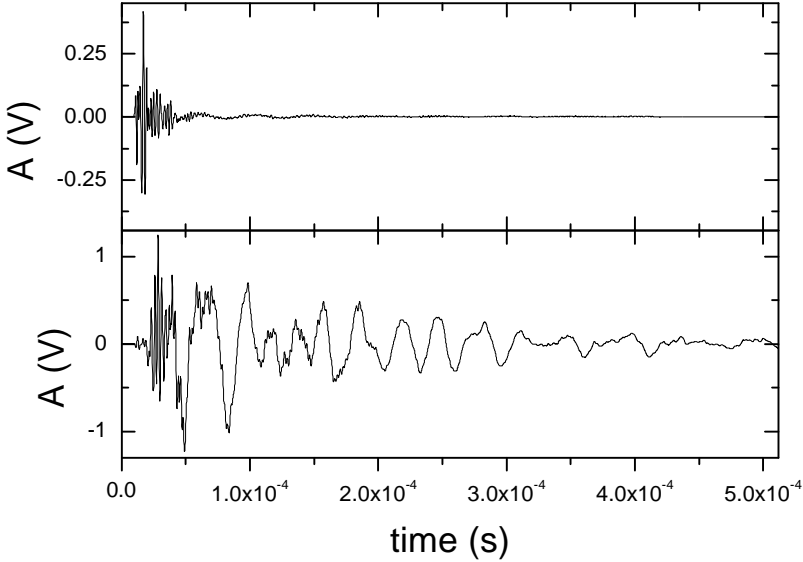

**Figure 5. The two main types of recorded AE waveforms. Top, type 1 pattern: AE waveform recorded in the stage of easy basal glide of the test a3, and assumed to correspond to a dislocation glide avalanche. Bottom, type 2 pattern: AE waveform recorded in the subsequent stage of the test a3, and assumed to correspond to twinning.**

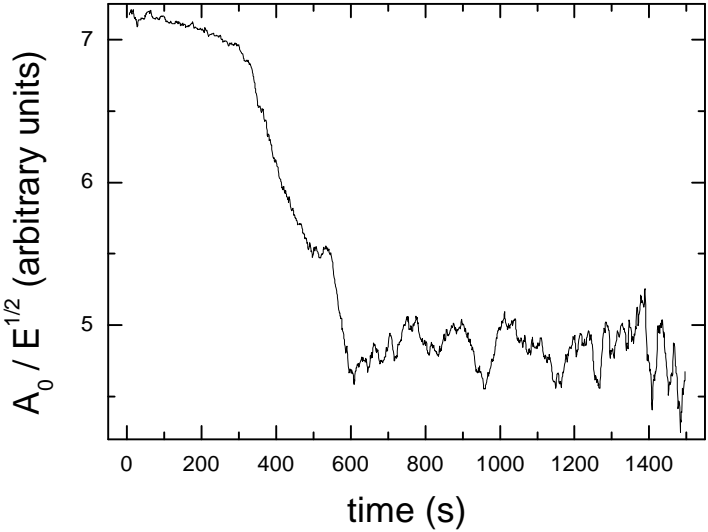

**Figure 6. Time evolution of the averaged ratio between AE event maximum amplitude and square root of AE event energy during the test a3.**

In addition to power law distributions of avalanche sizes, the complexity of dislocation dynamics in ice is expressed by complex time patterning and interactions [6]. A similar analysis, as described in [6], was performed for each test. A time clustering of AE events, i.e. some aftershock activity, is systematically observed. A deeper analysis can even be performed, if one looks at the S-type aftershocks triggered by a T-type mainshock (and vice-versa). This kind of analysis was performed for sample a3, in a region (time > 800s) where the proportion of T-type events is nearly constant. The average rate $n_T^S(t)$ of S-type events recorded after a T-type mainshock of any amplitude, per mainshock and per unit time, was calculated (Fig. 7a). The physical triggering of AE events is clearly identified at small time scales, as $n_T^S(t)$ is larger than its background value due to uncorrelated events at longer times. This observation cannot be an artefact induced by echoes. Indeed, because of the sample dimensions, echoes due to waves reflected on the sample surfaces are expected to occur on much shorter time scale (~ 5 µs) than the set dead time (120 µs) between two AE events. This analysis shows that a T-type event can trigger an S-type event. Conversely, figure 7b shows that an S-type event can trigger a T-type event. Some additional analysis (not detailed here) shows that the aftershock triggering of figures 7a and 7b cannot be the mere consequence of the self-induced triggering of slip events still present in T-type population and of twinning events still present in S-type population. Hence, this analysis illustrates the mutual interactions occurring between dislocation slip and twinning. The experimental values of $n_T^S(t)$ and $n_S^T(t)$ show evidence of the triggering of one process by the other one.

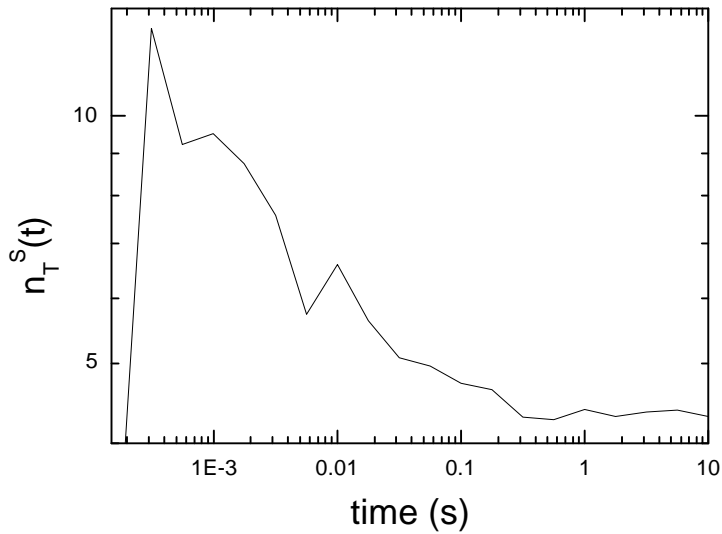

**Figure 7a. Average rate $n_T^S(t)$ of S-type events recorded after a T-type event of any amplitude, per mainshock and per unit time, during the test a3 and test time > 800s.**

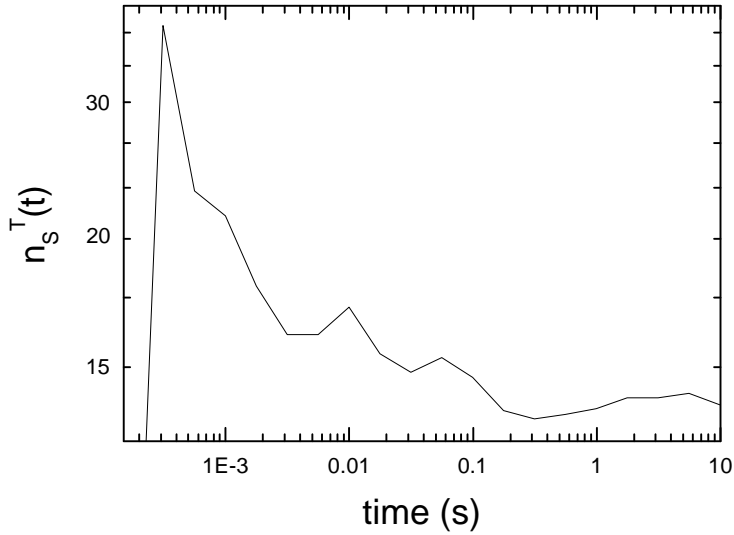

**Figure 7b.** Average rate $n_S^T(t)$ of T-type events recorded after a S-type event of any amplitude, per mainshock and per unit time, during the test a3 and test time > 800s.

## Discussion and conclusion

We report here AE experiments during plastic deformation of hcp metallic single crystals. Two sources of AE are identified: dislocation glide avalanches and twinning. These two sources seem to correspond to two different types of AE waveforms (Fig. 5). From the analysis of the stress-strain curves and from arguments, developed in the literature, concerning specific properties of AE related to slip or twinning, the global AE events population can be separated into two populations, characterized by a small and a large proportion of twinning. Both populations show the same power law distribution for the pdf of their AE energies (Fig. 4). In addition, through the analysis of time correlations, mutual interactions between twinning and slip are highlighted (Fig. 7). Evidence is given of the triggering of one process by the other one. This picture is physically sound: indeed, avalanches of gliding dislocations induce stress rearrangements which may trigger twinning events. Conversely, by changing locally the orientation of the crystal and consequently the local stress field, twinning may trigger dislocation glide avalanches.

The power law distribution of induced AE energies is the same and it is not important, whether dislocations are involved in a slip or in a twinning event. This pattern is also recurrent whatever the material composition (Cd or Zn-0.08%Al) or the orientation of the basal planes. In ice polycrystals [22], the propagation of avalanches was hindered by grain boundaries and a finite (grain)-size effect on avalanche size distributions was observed. This effect was characterized by a strong cut-off of the power law scaling towards large amplitudes as well as by an apparent change of the power law exponent. In the present experiments, undisturbed power law distributions are observed even when single crystals are highly twinned, thus revealing a difference of behaviour, with respect to dislocation dynamics, between twin and grain boundaries (at least in the system studied here). The values found for the exponent $t_E$ of the power law distributions of energies are in very good agreement with those found during creep experiments on ice single crystals, $t_E$ = 1.5±0.1. This result strongly suggests that the general features of the observed intermittent flow regime are of generic nature in plastic deformation of crystalline materials. The non-proportionality observed between AE activity and macroscopic deformation (Fig. 2) suggests that the emergence of the scale-free behaviour of plastic instabilities may not be related to the local deformation achieved by each instability but rather to the energy dissipated by it. Besides, it is worth underlining that identical power law distributions are found from two deformation modes: creep and strain controlled tests. One could wonder if stress relaxations occurring during strain controlled tests might prevent the emergence of the self-organized critical dynamics. Although a slight cut-off of the AE energy distributions cannot be excluded, the present results argue against this scenario.

As observed in ice single crystals, plastic instabilities in hcp metallic single crystals can be characterized by strong intermittency (Fig. 2), power law distributions of dissipated energy (Fig. 3), time correlations and aftershock triggering (Fig. 7). This scale-free pattern is independent of the type of material composition, of the nature of dislocations (i.e., those related to slip or to twinning) and whether the material exhibits strong forest hardening (i.e. multi-slip) or not. This evidence is new and strong argument in favour of a general, SOC-type, framework for dislocation-related plasticity. It is worth noting that a similar exponent is also found in microfracturation processes [23]. Such processes are significantly different of the present ones. Actually, they involve successive damage events that eventually lead to failure, whereas in our case AE results from stress relaxation events during a marginally stable state, without any cumulative effect. However, this intriguing similarity, that would require further attention, suggests that such an exponent may be a universal signature of stress relaxation processes.

## Acknowledgements

This work is a part of the Research Plan MSM 0021620834 that is financed by the Ministry of Education, Youth and Sports of the Czech Republic. The work received also support from the French Ministry of Foreign Affairs, and the Ministry of Education, Youth and Sports of the Czech Republic within the framework of the mobility program Barrande Nr. 2005-06-032-1.